\documentclass[runningheads]{llncs}
\usepackage[T1]{fontenc}
\usepackage{graphicx}
\usepackage{booktabs}
\usepackage[misc]{ifsym}
\newcommand{\corr}{(\Letter)}

\usepackage{xcolor} 
\usepackage{amsfonts} 
\usepackage{amsmath}

\newif\ifNota
\Notafalse

\newcommand{\datasetname}{Acquirers\ }

\begin{document}

\title{Evaluating Transfer Learning Methods\\on Real-World Data Streams:\\A Case Study in Financial Fraud Detection}

\titlerunning{Evaluating TL Methods on Real-World Data Streams}

\author{Ricardo Ribeiro Pereira\inst{1,2} \corr \and
Jacopo Bono\inst{1} \and
Hugo Ferreira\inst{1} \and \\
Pedro Ribeiro\inst{2} \and
Carlos Soares\inst{2} \and
Pedro Bizarro\inst{1}
}

\authorrunning{R.R. Pereira et al.}

\institute{
Feedzai, Portugal\\
\and
University of Porto, Portugal\\
\email{ricardo.ribeiro@feedzai.com}
}

\maketitle

\begin{abstract} 

When the available data for a target domain is limited, transfer learning (TL) methods can be used to develop models on related data-rich domains, before deploying them on the target domain.
However, these TL methods are typically designed with specific, static assumptions on the amount of available labeled and unlabeled target data.
This is in contrast with many real world applications, where the availability of data and corresponding labels varies over time.
Since the evaluation of the TL methods is typically also performed under the same static data availability assumptions, this would lead to unrealistic expectations concerning their performance in real world settings.
To support a more realistic evaluation and comparison of TL algorithms and models, we propose a data manipulation framework that (1) simulates varying data availability scenarios over time, (2)\ creates multiple domains through resampling of a given dataset and (3)\ introduces inter-domain variability by applying realistic domain transformations, e.g., creating a variety of potentially time-dependent covariate and concept shifts.
These capabilities enable simulation of a large number of realistic variants of the experiments, in turn providing more information about the potential behavior of algorithms when deployed in dynamic settings.
We demonstrate the usefulness of the proposed framework by performing a case study on a proprietary real-world suite of card payment datasets.
Given the confidential nature of the case study, we also illustrate the use of the framework on the publicly available Bank Account Fraud (BAF) dataset.
By providing a methodology for evaluating TL methods over time and in realistic data availability scenarios, our framework facilitates understanding of the behavior of models and algorithms.
This finally leads to better decision making when deploying models for new domains in real-world environments.

\end{abstract}

\section{Introduction}
\label{sec:introduction}

Machine learning (ML) models often require large volumes of labeled data to achieve strong predictive performance.
However, in many real-world applications, obtaining sufficient labeled data can be difficult and costly.
Transfer learning (TL) addresses this challenge by leveraging knowledge from one or more source domains to improve performance on a target domain with limited data.
Most TL methods and evaluation protocols assume fixed conditions regarding the availability of labeled and unlabeled data, such as having a large labeled source dataset and only unlabeled target data.
However, in many real-world industry settings, these conditions are not permanent, as data is progressively collected and labeled over time.

One example of this setting is financial fraud detection.
This task involves monitoring streams of financial transactions from different financial institutions, \emph{domains} in the TL terminology, and classifying each transaction as fraudulent or legitimate.
New institutions may initially lack historical data, but, typically, the volume of financial transactions quickly increases over time.
However, labeling a transaction as fraudulent often depends on customer complaints and/or manual reviews by analysts, leading to a delay between the moment a transaction is recorded and when it is labeled.
This delay can range from several days to a few months, affecting the training and evaluation of ML models.
While TL can in principle help mitigate the issues of having insufficient data at the onset, and insufficient labeled data at a later stage, the evolving nature of the data availability itself presents an additional challenge.
TL methods are designed for fixed conditions and their performance is expected to change significantly when those conditions are violated. However, they are typically evaluated under those fixed (and favorable) conditions, which would lead to unrealistic expectations concerning their performance in real world settings.
The problem therefore remains on how to evaluate TL methods in a way that reflects the real-world dynamic data constraints, such as those encountered in fraud detection.

To address this challenge, we propose an evaluation framework that simulates the dynamic nature of real-world data streams.
Our framework provides three key capabilities:
(1) creating multiple domains from a given dataset through resampling, enabling systematic TL evaluation even when few datasets are available;
(2) applying transformations to the data, hence reproducing realistic data shifts over time and across domains;
and (3) simulating the gradual arrival of data and labels over time, mimicking the evolving nature of industry environments.
These combined features enable our framework to systematically generate a large number of experiments, making it possible to assess TL methods across a wide range of realistic scenarios.

We perform a case study using our framework on a suite of proprietary real-world datasets containing payment events from multiple financial institutions.
This case study demonstrates how insights derived from our evaluation framework can inform practical decisions, such as model selection, deployment timing, and the prioritization of data collection efforts. Given the confidential nature of the case study dataset, we perform a similar analysis on the publicly available Bank Account Fraud (BAF) dataset.
The source code that implements the evaluation framework, along with the configurations used for the experiments on the public dataset, will be available at \url{https://github.com/feedzai/tred} after publication.

The remainder of this paper is structured as follows:
Section~\ref{sec:relatedwork} formalizes our problem setting and compares it with traditional TL setups studied in academia;
Section~\ref{sec:method} introduces the design of our evaluation framework and its key components;
Section~\ref{sec:experimental_setup} describes how we apply the framework in practice, detailing the datasets, experimental setup, and TL methods evaluated;
Section~\ref{sec:results} presents the results and their practical implications in an industry setting;
and Section~\ref{sec:conclusion} summarizes our contributions and highlights the broader impact of our work.

\section{Background and related work}
\label{sec:relatedwork}

In this section, we formalize the problem setting and introduce the notation used throughout the paper (Section~\ref{subsec:problem_definition}).
We then review traditional TL paradigms, highlighting their assumptions and differences from our use case (Section~\ref{subsec:tl_paradigms}).
Finally, we discuss common evaluation strategies for TL and motivate the need for a new framework that better captures real-world data dynamics (Section~\ref{subsec:tl_benchmarks}).

\subsection{Problem definition}
\label{subsec:problem_definition}

We consider the machine learning setting where data is collected from multiple domains over time, with labels becoming available after a delay.
This is a common scenario in many real-world applications, such as fraud detection, where instances (e.g., transactions) are initially unlabeled and only later confirmed as fraudulent or legitimate.
To formalize this problem, we assume there are $m$ source domains $\mathcal{D}_{S_1},\ldots,\mathcal{D}_{S_m}$ and a target domain $\mathcal{D}_T$.
Each domain $\mathcal{D}_d$ (including the target) is associated with a dataset
$D_d=\{(x_i,y_i,t^x_i,t^y_i) \mid i=1,\ldots,n_d\},$
where $x_i \in \mathcal{X}_d$ is a feature vector, $y_i \in \mathcal{Y}_d$ is the label, $t^x_i$ is the timestamp when $x_i$ is collected, and $t^y_i \ge t^x_i$ is the timestamp when $y_i$ becomes available.\footnote{To simplify notation, we will sometimes use the letter $i$ to index the entries of the dataset without explicitly stating $i=1,\ldots,n_d$.}
At any given time $t$, $D_d$ can be decomposed into a labeled dataset $D^L_d(t) = \{ (x_i, y_i) \mid t^y_i \leq t \}$ which consists of all instances that have already received their labels by time $t$, and an unlabeled dataset $D^U_d(t) = \{ x_i \mid t^x_i \leq t < t^y_i \}$ which consists of instances that have been observed but their labels are still unavailable at time $t$.

Eventually, at some time $t_a$, the target domain $\mathcal{D}_T$ is introduced, initially without any data ($D^L_T(t_a) = D^U_T(t_a) = \emptyset$), and target domain data and labels begin to be collected from that point on.
Our goal is to leverage $D_{S_1},\ldots,D_{S_m}$ and $D_T$ to learn a predictive function $f_T: \mathcal{X}_T \to \mathcal{Y}_T$ that approximates $P_T(Y|X)$.
Over time, as more data and labels become available, $f_T$ can be updated to improve its approximation of $P_T(y\mid x)$.

\subsection{Transfer learning paradigms}
\label{subsec:tl_paradigms}

Different TL paradigms have been explored, each making different assumptions about the datasets used to train the ML models.
All these paradigms assume that a large volume of labeled data is available from the source domains.
Their main difference relates to the available target domain data at training time.

In Domain Generalization (\textbf{DG})~\cite{zhou2022domain,wang2022generalizing}, the goal is to use the source domain datasets to learn a predictive function $f$ that generalizes to the target domain without access to any data from $\mathcal{D}_T$.
As such, at training time, $D^L_T = D^U_T = \emptyset$.
In Unsupervised Domain Adaptation (\textbf{UDA})~\cite{wilson2020survey}, in addition to the source domain datasets, there is an unlabeled target domain dataset that can be used to adapt the predictive function $f_T$ to the target domain $\mathcal{D}_T$.
This means that, at training time, $|D^U_T|>0$ while $D^L_T = \emptyset$.
In Supervised Domain Adaptation (\textbf{SDA})~\cite{wang2018deep}, in addition to the source domain datasets, there is both a large unlabeled dataset and a small labeled dataset from the target domain, which are used to adapt the predictive function $f_T$ to the target domain $\mathcal{D}_T$.
As such, at training time, $|D^U_T|\gg|D^L_T|>0$.
In Multi-Domain Learning (\textbf{MDL})~\cite{yang2014unified}, the goal is to use datasets from multiple domains to learn a single predictive function $f$ that performs well across all observed domains simultaneously.
Here, at training time, labeled data is available from all domains, i.e., $\forall d, |D^L_d|\gg 0$.

Each of these paradigms operates under specific assumptions about data availability, but none of them account for the progressive collection of data and possible label delay.
In contrast, our problem setting requires a framework that can systematically model the evolving availability of data and labels over time.

\subsection{Evaluation of TL methods}
\label{subsec:tl_benchmarks}

Various datasets have been used to evaluate TL methods, under the different paradigms discussed in the previous section.
Most TL benchmarks focus on image classification, including datasets such as Office-31~\cite{saenko2010adapting}, Office-Caltech10~\cite{gong2012geodesic}, Office-Home~\cite{venkateswara2017deep}, DomainNet~\cite{peng2019moment}, and PACS~\cite{li2017deeper}.
Beyond computer vision, the Amazon Reviews dataset~\cite{blitzer2007biographies} is often used for sentiment analysis.

Another common strategy is to evaluate TL methods across different datasets of the same task.
Examples include: digit classification (MNIST~\cite{lecun1998gradient}, USPS~\cite{hull1994database}, SVHN~\cite{netzer2011reading}); large-scale image recognition (ImageNet~\cite{deng2009imagenet}, CiFAR~\cite{krizhevsky2009learning}, Caltech~\cite{griffin2007caltech}); and semantic segmentation (GTA5~\cite{richter2016playing}, CityScapes~\cite{cordts2016cityscapes}).

Additionaly, some tools have been developed to facilitate the evaluation of TL methods in specific fields.
One example is DomainATM~\cite{guan2023domainatm}, an open-source MATLAB package for domain adaptation in medical data analysis.
It provides dataset management functionalities, visualization tools, and a collection of domain adaptation methods with built-in evaluation capabilities.

However, both DomainATM and traditional TL benchmarks assume a static evaluation setting, where data availability conditions remain fixed.
This assumption overlooks the temporal dynamics present in real-world applications, such as fraud detection, where data and labels arrive progressively over time.
As a result, existing evaluation strategies are insufficient for assessing TL methods in dynamic environments.
Addressing this gap requires a framework that systematically models the evolving availability of data and labels, enabling more realistic evaluations that reflect real-world deployment scenarios.

\section{Method}
\label{sec:method}

\begin{figure}[t]
    \centering
    \includegraphics[width=0.9\linewidth]{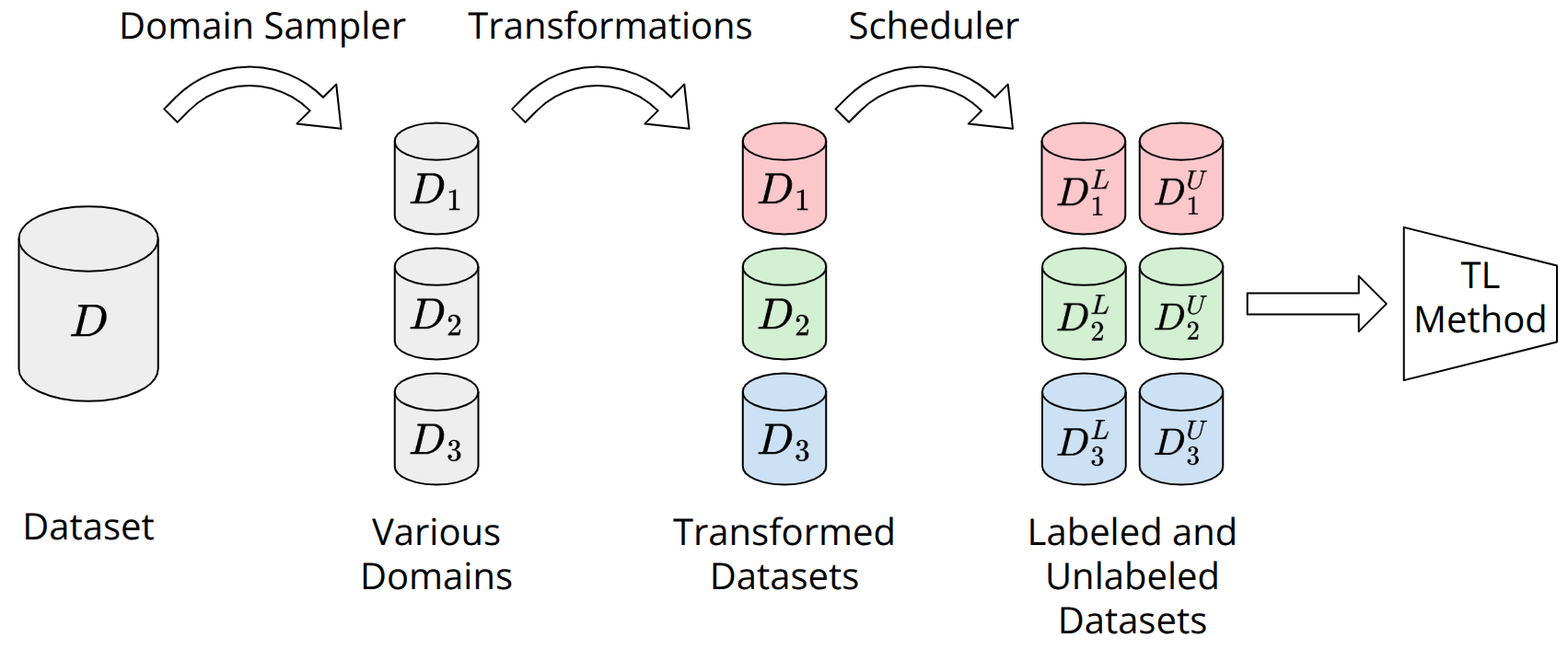}
    \caption{The evaluation framework is composed of three sequential components: domain sampler, transformations, and scheduler. The number of domains depicted in the diagram is just an example.}
    \label{fig:framework_diagram}
\end{figure}

Our evaluation framework consists of three components (Figure~\ref{fig:framework_diagram}).
First, the \textit{domain sampler} builds multiple domains from a single dataset, enabling systematic TL evaluation even when few datasets are available.
Next, the \textit{transformations} introduce controlled variations to each domain, to reproduce real-world data shifts over time and across domains.
Finally, the \textit{scheduler} simulates the progressive arrival of data and labels, enabling the evaluation of TL methods under diverse data availability conditions.
We describe each component with more detail in the following subsections.

\subsection{Domain Sampler}

\begin{figure}[t]
    \centering
    \includegraphics[width=0.9\linewidth]{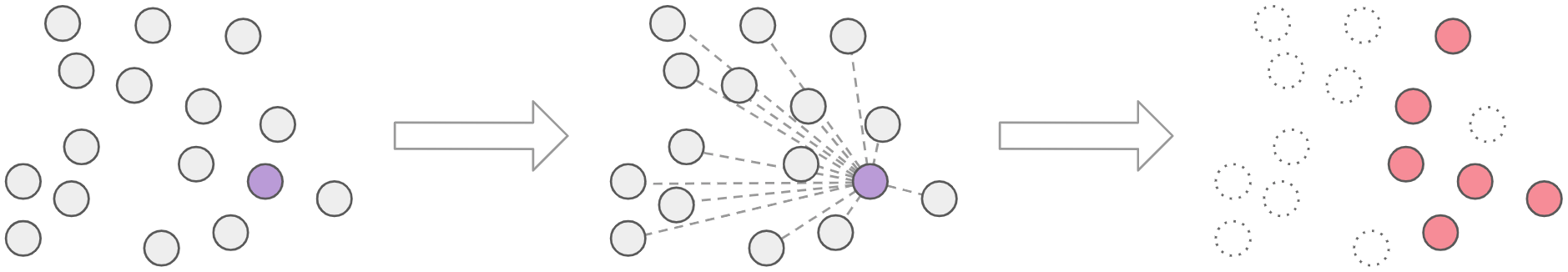}
    \caption{Toy example of sampling a domain from dataset. First, an anchor instance is selected (purple point). Then, the distances to all other instances are computed. Lastly, the instances are sampled with probability decreasing as distance increases.}
    \label{fig:domain_sampler}
\end{figure}

The \textit{domain sampler} creates domains from a dataset by randomly selecting anchor instances, followed by resampling the events according to the distance to these anchors, as illustrated in Figure~\ref{fig:domain_sampler}.

More formally, the \textit{domain sampler} is a stochastic process that receives a dataset $D=\{(x_i,y_i,t^x_i,t^y_i)\}$, a distance function $\delta$, a real number $\lambda$, and a positive integer $k$, and outputs a set of datasets $\{D_1,\ldots,D_{k}\}$ where $D_d \subseteq D$ for all $d\in\{1,\ldots,k\}$. 
To extract each $D_d$, the \textit{domain sampler} first selects an instance $x_{\text{anchor}}$.
Then, each instance $x_i$ is assigned a probability of being included in this domain, which decreases exponentially with its distance to $x_{\text{anchor}}$:
\[
P(x_i | x_{\text{anchor}}) = e^{-\lambda \delta(x_i,x_{\text{anchor}})} \, .
\]
The decay rate of the exponential is controlled by the scaling factor $\lambda$, which regulates the expected domain size. Finally, instances are sampled randomly according to their respective probability.

\subsection{Transformations}

The \textit{transformations} apply controlled modifications to the datasets.  These transformations are defined by the user to better suit their setting. For example, transformations that may make sense in the image domain would not be suitable for tabular data and vice-versa. 
Each \textit{transformation} should ideally be parameterized differently for each domain, provoking some level of domain shift.
Furthermore, they can be designed to depend on the timestamp of the instance, which effectively simulates data drift over time or seasonalities.

Each \textit{transformation} can be described as a function $\Phi_\theta: (x,y,t^x,t^y) \mapsto (x',y',t'^x,t'^y)$, parameterized by $\theta$.
This general formulation allows the instantiation of various types of changes, for example:
\begin{itemize}
    \item covariate shift (change in $P(X)$): $x'=\phi(x;\theta)$;
    \item concept shift (change in $P(Y|X)$): $y'=\phi(x,y;\theta)$;
    \item data drift (change in $P(X)$ over time): $x'=\phi(x,t^x;\theta)$.
\end{itemize}

\begin{figure}[t]
    \centering
    \includegraphics[width=0.9\linewidth]{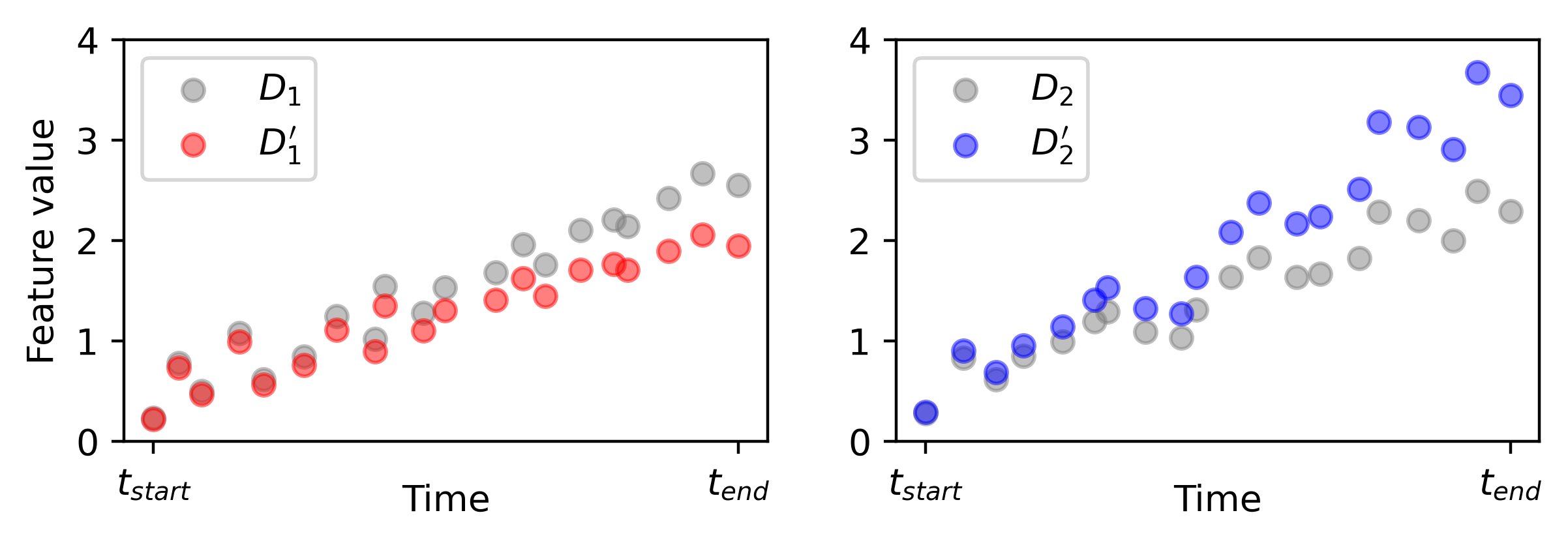}
    \caption{Toy example showing how applying the same transformation with different parameters affects the same feature of datasets from two domains.}
    \label{fig:toy_transform}
\end{figure}

If the transformations are parameterized differently for each experiment, the results will express a distribution of each methods' performance on related settings, increasing the robustness of the results.
We describe a set of \textit{transformations} for tabular data in detail in Section~\ref{subsec:exp_transf} (as well as making them available with our code) and provide a toy example in Figure~\ref{fig:toy_transform}.

\subsection{Scheduler}
\label{subsec:method_scheduler}

The \textit{scheduler} orchestrates two processes: (1) the progressive arrival of instances and labels over time and (2) the performance estimation over time.

The first process (progressive data arrival) is achieved by discretizing the time range of the target dataset in contiguous periods.
At each step, the test period advances, while the training set expands to include all data up to that point.
More formally, the \textit{scheduler} receives datasets $D_{S_1},\ldots,D_{S_m},D_T$ and a sequence of user-defined timestamps $t_1,\ldots,t_l$ s.t. $\min(t^x_i)\le t_1 <\ldots<t_l\le \max(t^x_i)$ for $t^x_i \in D_T$.
At each time step $t_a$ for $a=1,\ldots,l-1$, it decomposes all source and target domain datasets $D_d$ into $D^L_d(t_a)$ and $D^U_d(t_a)$, as described in Section~\ref{subsec:problem_definition}.

The second process (performance estimation) is achieved by leveraging the data splits that result from the first process to train the TL methods under study and evaluate them on the target domain instances s.t. $t_a\le t^x_i<t_{a+1}$.\footnote{Notice that the label delay is ignored for the purpose of evaluating the methods.}

Since this process is repeated for each time step $t_a$, every model is evaluated multiple times throughout the evolving target dataset.
This allows for an analysis of performance trends over time, highlighting how different TL methods adapt to increasing data availability.
Figure~\ref{fig:baf_splits} presents an example of this scheduling.

\section{Experimental Setup}
\label{sec:experimental_setup}

In the previous section, we introduce the general architecture of the framework, which is designed to be broadly applicable to many real-world scenarios.
In this section, we detail how we apply this framework to our specific use-case, including a description of the datasets that we use, the methods that we test and other design decisions that are specific to our experimental setup.\footnote{Implementation details and code will be available at \url{https://github.com/feedzai/tred} after publication.}

\subsection{Datasets}
\label{subsec:exp_datasets}

In this section, we provide details about the dataset used in our case study, but due to confidentiality constraints, we can only share general metrics.
The \datasetname is a real-world proprietary dataset, containing payment events from 4 different financial institutions (\emph{domains}) over a period of 41 weeks ($\approx9$ months).
Each domain has $\approx 5$M events, but their fraud rates (relative frequency of the positive class) varies between $\approx 0.01\%$ and $\approx 0.4\%$.
Each instance has 58 features (52 numerical and 6 categorical), the event timestamp and the fraud label.

Because the above dataset is confidential, we also perform experiments on the publicly available Bank Account Fraud (BAF) dataset~\cite{jesus2022turning}. BAF is a publicly available synthetic bank account fraud dataset.\footnote{In fact, the authors published 6 different variations of this dataset, but we just use the "Base" variant without device\_fraud\_count and device\_os features.}
It contains one million examples of account opening applications, some of which are fraudulent, from February through September.
Each instance has 28 features (24 numerical and 4 categorical), the time information and the label.

In both experiments, each numerical feature from each domain is standardized to have 0 mean and 1 standard deviation.
Also, each categorical feature is label encoded~\cite{borisov2022deep}, i.e. each category is mapped to an integer starting from 0, to enable the use of embedding layers.
Furthermore, to address class imbalance, we oversample the minority class during training by constructing batches with a fixed $10\%$ positive class ratio.
For evaluation, the original proportion is used.

\subsection{Domain Sampler}
\label{subsec:exp_ds}

The BAF dataset does not contain any explicit separation of domains.
As such, in our experiments, we use the \textit{domain sampler} to the create $4$ domains: $3$ sources and $1$ target.
Since the domains are randomly sampled, without loss of generality, we always select the first one to be the target.
Given that BAF is a tabular dataset, we define a distance function $\delta$ to compare rows containing a set of numerical features $\mathcal{N}$ and a set of categorical features $\mathcal{C}$.
For numerical features, we compute the squared difference between their standardized values.
For categorical features, we use an indicator function that returns 0 if the values are the same, and 1 otherwise.
\[
\delta(x_i, x_j) = \sum_{f \in \mathcal{N}} \left(\frac{x_{i,f}-x_{j,f}}{\sigma_f}\right)^2 + \sum_{f \in \mathcal{C}} \mathbb{I}[x_{i,f} \neq x_{j,f}] \, ,
\]

\subsection{Transformations}
\label{subsec:exp_transf}

We define three types of operations that can be applied to features of tabular datasets.
$\phi_1$ rescales numerical features by multiplying their values by factor.
$\phi_2$ computes a weighted average between the value of a numerical feature and a certain anchor value $\beta$.
$\phi_3$ resamples some values of a categorical feature, approximating its relative frequencies to some marginal distribution $P(X'_j)$.
\[
\begin{array}{l}
    \phi_1(x_{i,j}, t^x_i;\theta) = x_{i,j} \cdot \alpha^{\tau(t^x_i)}\text{, where } \theta=(\alpha,\tau)\\
    \phi_2(x_{i,j}, t^x_i;\theta) = (1-\gamma \cdot \tau(t^x_i)) \cdot x_{i,j} + (\gamma \cdot \tau(t^x_i))\cdot \beta \text{, where } \theta=(\beta,\gamma,\tau)\\
    \phi_3(x_{i,j}, t^x_i;\theta) \sim (1-\tau(t^x_i)) P(X_j) + \tau(t^x_i) P(X'_j)\text{, where } \theta=(P(X'_j),\tau)
\end{array}
\]
In all three \textit{transformations}, the parameter $\tau$ is a function that effectively controls the magnitude of the transformation, given the timestamp of the instance.
We used three versions of $\tau$ (not in a one-to-one correspondence with transformations):
(1) a constant function equal to $1$, simulating fixed changes between domains, such as changes of currency;
(2) a linear function going from $0$ at the start of a dataset to $1$ at the end, simulating domains that are gradually drifting over time, such as countries having different levels of inflation;
(3) a sine wave with a parameterized period, simulating seasonal behaviors, such as higher or lower spending on the weekends.

\subsection{Scheduler}
\label{subsec:exp_scheduler}

For each experiment, given a set of source and target datasets with time span $[t_s, t_e)$, we define $t_\alpha$ and $t_\beta$ as the start of the data availability from the source and target domains respectively, and $t_\gamma$ as the end of the experiment such that $t_s \le t_\alpha < t_\beta < t\gamma \le t_e$.
The target domain data in the interval $[t_\alpha, t_\beta]$ is ignored to ensure that the first training split contains only source domain data, mimicking real-world deployment scenarios where historical target data is unavailable at launch.
We also define the time interval between model updates $\Delta_t$, which also defines the duration of each test split.
Lastly, since neither dataset contains a label timestamp, we define a fixed label delay $\Delta_l$ such that $t^y_i = t^x_i + \Delta_l$.
Using these parameters, the \textit{scheduler} simulates the progressive arrival of data as described in Section~\ref{subsec:method_scheduler}, generating a sequence of timestamps $t_1,\ldots,t_l$ s.t.
\[
t_1=t_\beta, \quad t_{a+1}=t_a+\Delta_t, \text{ for } a=1,\ldots,l-1
\]
where $t_l$ it the largest timestamp satisfying $t_l \le t_\gamma$.

For the \datasetname dataset, we use the time unit of one week, with timestamps indexed in the range $[0,41)$, and setting $\Delta_t =2$ and $\Delta_l=4$.
In each experiment, $t_\alpha$ is randomly selected from $\{0,\ldots,7\}$ to introduce variability while ensuring the framework leverages the entire data range.
Then, $t_\beta$ is set as $t_\alpha+16$, ensuring 16 weeks of available source domain data before the target appears, and $t_\gamma$ is set as $t_\alpha+34$, resulting in 9 contiguous test periods.

\begin{figure}[t]
    \centering
    \includegraphics[width=0.9\linewidth]{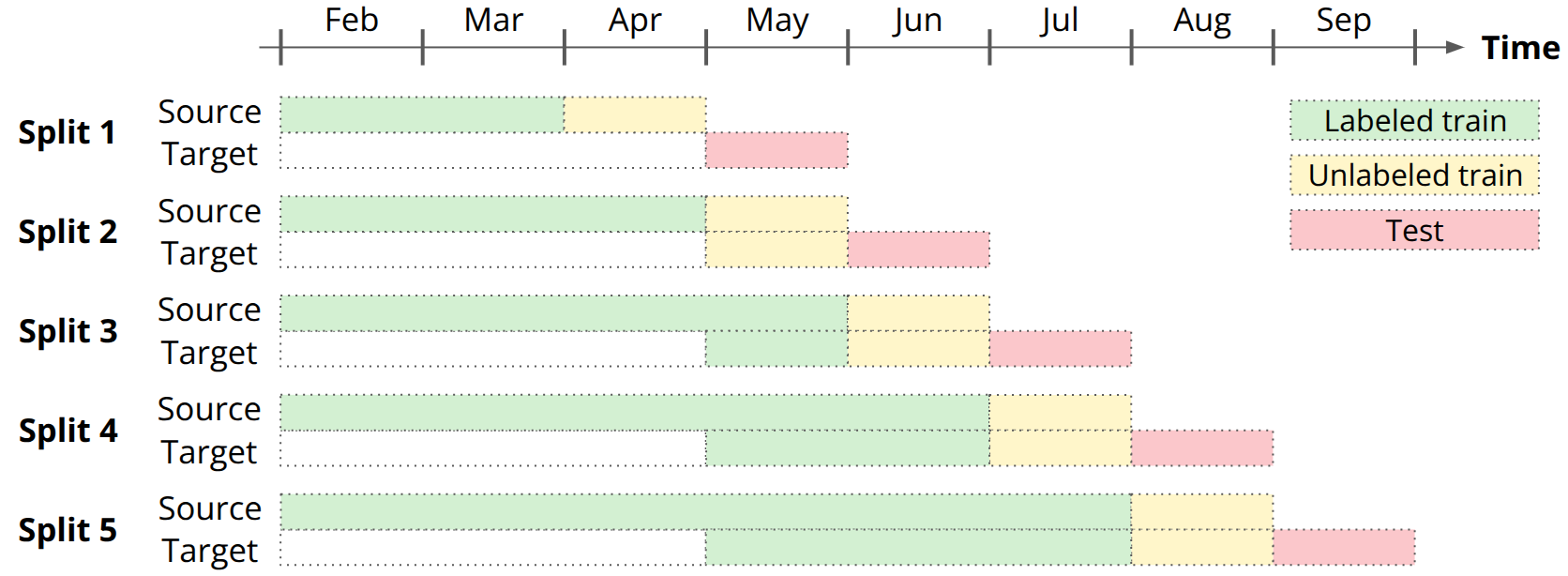}
    \caption{Schedule of the splits we used for the BAF experiments.}
    \label{fig:baf_splits}
\end{figure}

For the BAF dataset, we use the time unit of one month with timestamps indexed in the range $[0,8)$.
We set $t_\alpha=0$, $t_\beta=3$, $t_\gamma=8$ and $\Delta_t = \Delta_l = 1$.
The resulting schedule for this dataset is depicted in Figure~\ref{fig:baf_splits}.

\subsection{TL Methods}
\label{subsec:exp_methods}

Given that our goal is to illustrate how our framework supports the comparison of TL algorithms under varying data availability conditions and not to benchmark particular TL methods exhaustively, we implemented and tested one representative method from each of the four TL paradigms discussed in Section~\ref{subsec:tl_paradigms}: Multi-Task Autoencoder (\textbf{MTAE})~\cite{ghifary2015domain} for DG, Domain Adaptation Neural Networks (\textbf{DANN})~\cite{ganin2016domain} for UDA, Minimax Entropy (\textbf{MME})~\cite{saito2019semi} for SDA, and Multinomial Adversarial Networks (\textbf{MAN})~\cite{chen2018multinomial} for MDL.
Additionally, we tested three MLP baselines, differing only in their training data: \textbf{BL-S} is trained only with labeled source domain data; \textbf{BL-T} is trained only with labeled target domain data; and \textbf{BL-A} is trained using all labeled data available.

\subsection{Evaluation}
\label{subsec:exp_evaluation}

For all methods, we use the latest $30\%$ of the labeled training data from each domain as a holdout validation set for early stopping.
The stopping criterion is based on the average predicted performance across all domains, measured as Recall at $1\%$ false positive rate (FPR), which is a standard evaluation metric used in fraud detection tasks.
This ensures a consistent stopping strategy, even when labels are scarce.

For each experiment, we compute paired t-tests for each pair of methods at every data split, to assess the statistical significance of the observed performance differences.
Given the substantial number of comparisons, we controlled the False Discovery Rate (FDR) at $1\%$ using the Benjamini–Hochberg procedure, which reduces the risk of identifying spurious effects.

\subsection{Pre-training and hyperparameter tuning}
\label{subsec:exp_hpt}

Many TL methods use pre-trained state-of-the-art models to initialize the parameters of their deep learning components.
In our experiments, we pre-train an MLP-based autoencoder using the first three months of source domain data, using a typical encoder-decoder architecture with reconstruction loss defined per feature type: we use mean squared error for numerical features (after standardization) and cross-entropy loss for categorical features.
This self-supervised learning phase enables the networks to learn robust feature representations before applying specific transfer learning methods.

To optimize the autoencoder architecture, we conduct a hyperparameter search over 200 randomly sampled configurations.
The search space includes variations in the number and size of hidden layers, the size of the latent space, the learning rate, regularization techniques (dropout, normalization), and the inclusion of skip connections.
For method-specific hyperparameters, we primarily followed the values recommended in the respective papers.
The details of the search space and best hyperparameters are provided in the code repository.

The encoder block of the best-performing autoencoder, selected based on validation loss, is then used to initialize the feature extractors of the TL methods.
For their classifier components, we used a simple architecture with a single hidden layer followed by the output layer.

\section{Results}
\label{sec:results}

In this section, we first present the results from our case study on the proprietary \datasetname dataset, after which we discuss the results on the publicly available BAF dataset.
Finally, we discuss the practical implications of these findings and describe how industry practitioners could use them to guide their decision-making process.

\subsection{\datasetname dataset case study}
\label{subsec:results_rma}

\begin{figure}[t]
    \centering
    \includegraphics[width=0.9\linewidth]{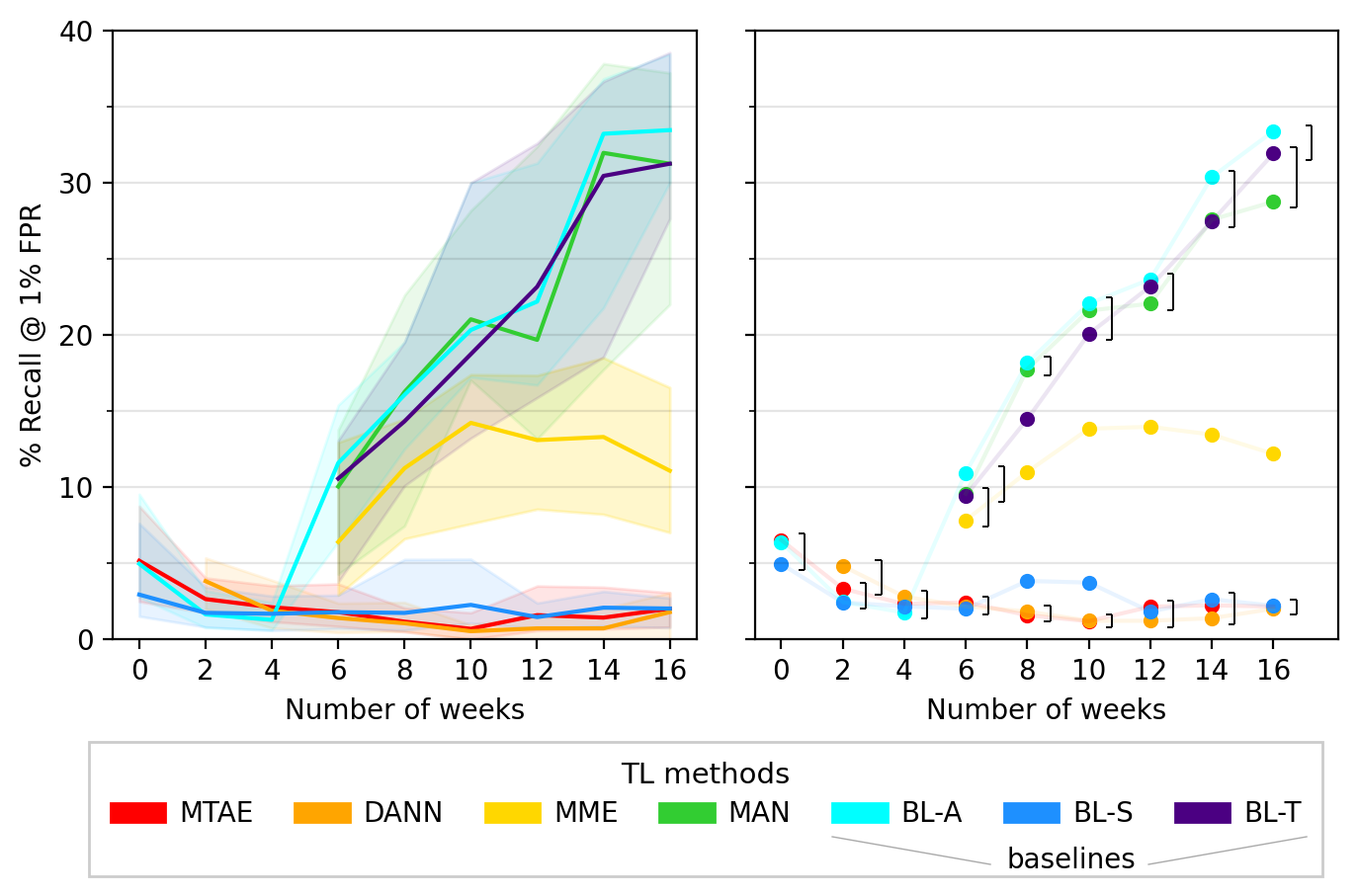}
    \caption{Predictive performance (recall at $1\%$ FPR) of each method over time on the \datasetname dataset.
    The left panel shows the median recall per method across experiments (solid lines) and their interquartile ranges (shaded).
    The right panel presents statistical comparisons at each time step, where each point represents the average recall across experiments, and brackets group methods that are not significantly different, after correcting for multiple comparisons.}
    \label{fig:rma_res}
\end{figure}

We conducted $48$ experiments on the \datasetname dataset, following the schedule described in Section~\ref{subsec:exp_scheduler}.
Figure~\ref{fig:rma_res} depicts the results of these experiments, showing the evolution of predictive performance over time for various baselines and TL methods.
The x-axis represents the time elapsed since the target domain appeared, while the y-axis depicts the recall percentage at $1\%$ FPR, which is a standard evaluation metric for the fraud detection problem.

There is a clear distinction between methods that leverage labeled target domain data and those that do not.
As such, we identify three groups of methods:
\begin{itemize}
    \item MTAE, DANN and BL-S, which don't use any target domain labels to train, maintain relatively stable performance throughout, but are consistently surpassed by the other methods.
    \item MAN, BL-A and BL-T, despite requiring target labels before their initial deployment, immediately outperform the other methods, and continue to improvement with subsequent increases in data availability, with an average gain of approximately 4 percentage points of recall per model update;
    \item the MME method initially follows the trend of the previous group, but its performance plateaus earlier, suggesting that its semi-supervised approach is more beneficial when labeled target data is scarce, but offering diminishing returns as more labels become available;
\end{itemize}
The statistical tests confirm that, in most splits, performance differences within each of the identified groups are not significant, while differences between groups typically are.

\subsection{BAF dataset}
\label{subsec:results_baf}

\begin{figure}[t]
    \centering
    \includegraphics[width=0.9\linewidth]{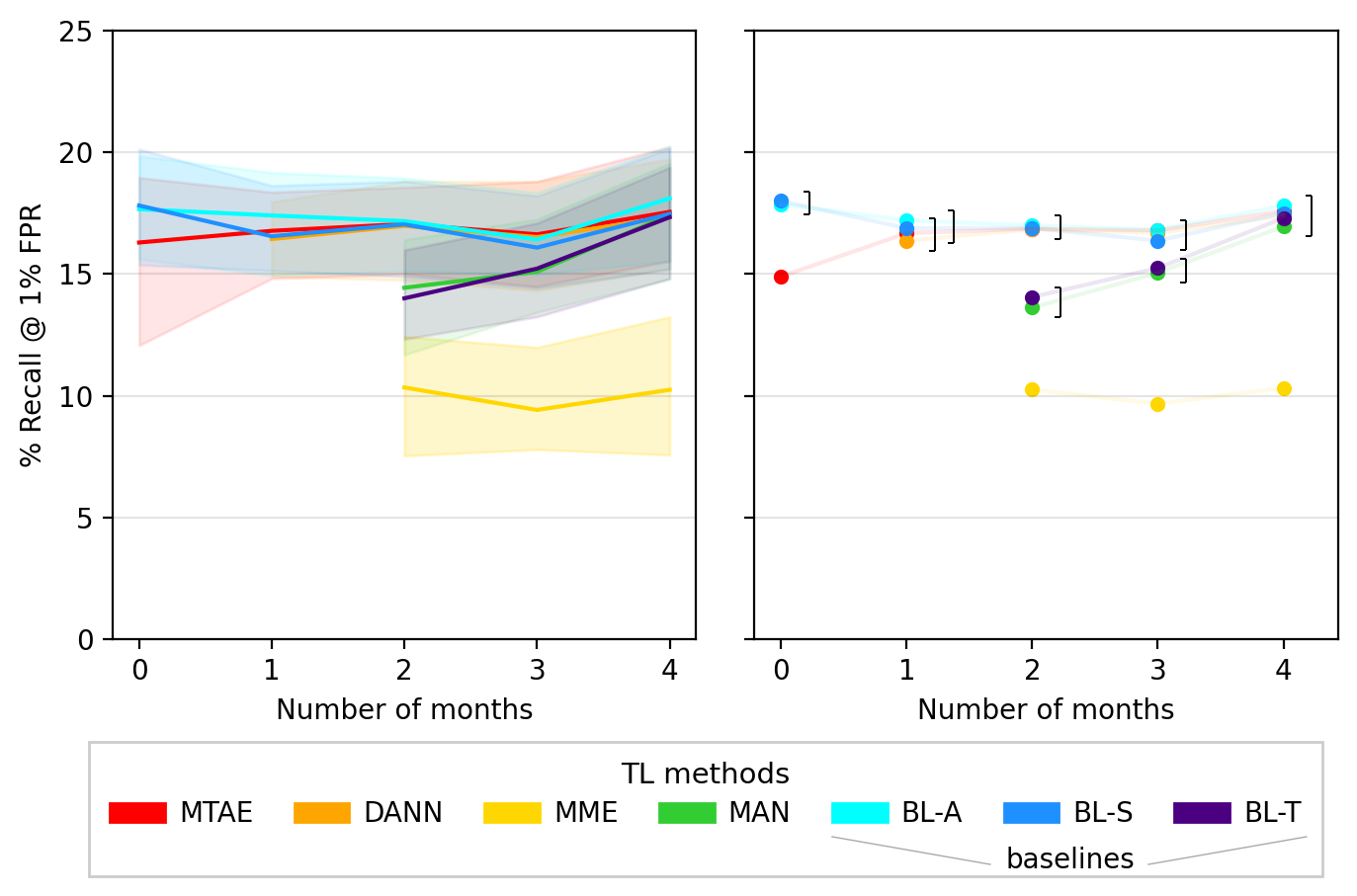}
    \caption{Predictive performance (recall at $1\%$ FPR) of each method over time on the BAF dataset.
    The left panel shows the median recall per method across experiments (solid lines) and their interquartile ranges (shaded).
    The right panel presents statistical comparisons at each time step, where each point represents the average recall across experiments, and brackets group methods that are not significantly different, after correcting for multiple comparisons.}
    \label{fig:baf_res}
\end{figure}

We conducted $128$ independent experiments on the BAF dataset.
In each experiment, we sampled four domains from the dataset, applied domain transformations (described in Section~\ref{subsec:exp_transf}), and followed the schedule depicted in Figure~\ref{fig:baf_splits} to train and evaluate the methods.
Figure~\ref{fig:baf_res} depicts the results of these experiments, in the same format of Figure~\ref{fig:rma_res}.

Similar to the previous experiments, the statistical tests allow us to identify three groups of methods:
\begin{itemize}
    \item MTAE, DANN, BL-A and BL-S show similar levels of recall, maintaining a stable distribution of predictive performance over time.
    This suggests that there is a limited benefit from the additional target domain data and labels.
    \item MAN and BL-T begin to perform significantly worse than the previous group of methods, but they improve steadily over time (gaining on average 2 percentage points of recall per model update) and eventually reaching the same level of performance.
    This improvement is not surprising, since both methods use exclusively labeled data from the target domain to train their classifiers.
    \item MME maintains a relatively stable performance throughout, but is consistently surpassed by the other methods.
\end{itemize}

Furthermore, we observe that the performance of methods such as MTAE and BL-S, which do not use any target data during training, is similar to the performance of BL-T, which follows the traditional ML approach of only using in-domain data to train.
This suggests that the source and target domains in the BAF dataset are relatively similar, which means that there is great potential for sharing knowledge across domains.

\subsection{Practical implications}
\label{subsec:results_practical}

Our experimental results highlight how the performance of different TL methods is affected by the evolving data availability conditions.
In general, methods that do not require target domain labels maintain stable but potentially limited performance, while methods that leverage target labels tend to noticeably improve as more data becomes available.
However, the extent and timing of this improvement vary across settings, meaning that both the effectiveness of these methods and the value of target labels depend on the specific characteristics of the dataset.
Recognizing these trends is essential for guiding real-world deployment decisions, helping practitioners select methods that align with their specific constraints and assess the cost-benefit of target labeling efforts.

For example, as seen in our \datasetname case study, practitioners may observe a clear performance gap between methods that leverage labeled target data and those that do not.
This gap suggests significant domain drift, making it difficult to share knowledge between domains.
In such cases, ML practitioners may decide to conduct a thorough exploratory data analysis to detect potential issues in the data collection pipeline.
They may also explore data pre-processing techniques, such as feature normalization, to mitigate domain discrepancies.
If the performance gap persists, deploying a DG-based solution initially can be a viable approach, but obtaining labeled target domain data should remain a priority to improve model performance.

Additionally, our results from \datasetname show that MDL methods (such as MAN and BL-A), which optimize performance across multiple domains simultaneously, perform comparably to domain-specific solutions.
In those cases, practitioners can benefit from these centralized solutions by minimizing the number of models that need to be developed and maintained.

Conversely, as seen in the BAF dataset experiments, practitioners may find that models trained exclusively on source domain data achieve performance levels similar to traditional in-domain models.
In those scenarios, DG methods can be confidently deployed for new domains without requiring immediate target domain labels, significantly reducing early labeling efforts and accelerating model deployment timelines.

\begin{figure}[t]
    \centering
    \includegraphics[width=\linewidth]{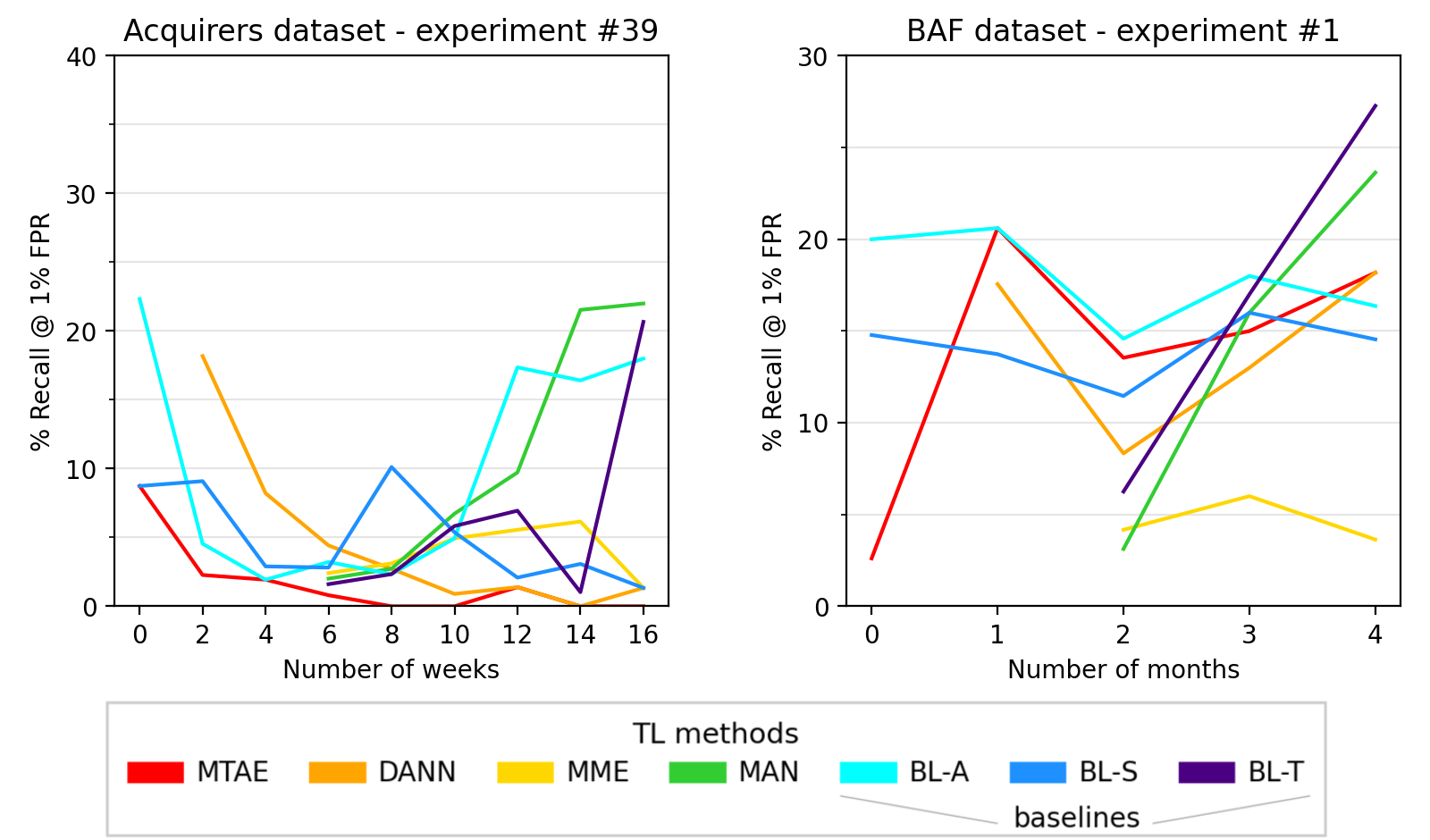}
    \caption{Predictive performance (recall at $1\%$ FPR) of each method over time.
    These panels show the results from two specific experiments (one from each dataset) that conflict with the general insights derived from the general analysis.}
    \label{fig:outliers}
\end{figure}

Finally, we highlight the importance of repeating multiple variants of the experiments to ensure robust conclusions.
Figure~\ref{fig:outliers} illustrates two specific cases that deviate from the general trends observed in our main analysis.
In the left panel, the results show that methods leveraging labeled target domain data, such as MAN and BL-A, only begin to outperform the other methods 12 weeks after the target domain appears, which is twice as long as observed in Section~\ref{subsec:results_rma}.
In the right panel, the results show that MAN and BL-T improve steeply over time, eventually surpassing all other methods, whereas the aggregated results in Section~\ref{subsec:results_baf} indicate no significant performance advantage for these methods.
These inconsistencies demonstrate the risk of drawing misleading conclusions from isolated experiments, which our evaluation framework helps to mitigate.

Beyond predictive performance, robustness over time is crucial for real-world deployment.
Some methods may maintain more stable performance, while others can exhibit a larger variance.
In high-risk applications such as fraud detection, consistency may be preferable to occasional peaks in performance.
Additionally, the trade-offs of other practical factors such as computationally efficiency, model update complexity, or interpretability, should be considered when selecting a TL method for real-world use.

\section{Conclusion}
\label{sec:conclusion}

In this paper, we introduce an evaluation framework designed to assess transfer learning methods under evolving data availability conditions.
Unlike traditional static benchmarks, our framework simulates the progressive arrival of data and labels, allowing for a more realistic and comprehensive evaluation of TL approaches in dynamic settings.
Additionally, by generating multiple realistic domain variations from the same dataset and applying controlled transformations, the framework enables systematic testing across diverse and realistic scenarios.
We demonstrate the capabilities of our framework through a case study on a proprietary dataset of card payment transactions, and perform an analogous study on the publicly available BAF dataset for reproducibility.
Our results illustrate how practitioners can leverage the framework to analyze TL performance trends over time, identify promising methods under varying data availability scenarios, and make informed decisions regarding deployment.
Thus, our framework improves upon traditional evaluations to address practical industry needs, providing a valuable tool for developing robust and adaptable machine learning solutions in dynamic real-world environments.

%

\end{document}